\documentclass[11pt]{article}
\usepackage{amsmath}
\usepackage{amssymb}
\usepackage{amscd}
\usepackage{enumerate}
\usepackage{mathdots}
\usepackage[latin1]{inputenc}
\newtheorem{theorem}{Theorem}
\newtheorem{corollary}{Corollary}
\newtheorem{lemma}{Lemma}

\newcommand{\CC}{\mathbb{C}}

\newcommand{\GG}{\mathbb{G}}
\newcommand{\RR}{\mathbb{R}}
\newcommand{\ZZ}{\mathbb{Z}}
\newcommand{\KK}{\mathbb{K}}
\newcommand{\MM}{\mathbb{M}}
\newcommand{\NN}{\mathbb{N}}

\newcommand{\HH}{\mathbb{H}}

\newcommand{\Ac}{\mathcal{A}}
\newcommand{\Bc}{\mathcal{B}}
\newcommand{\Ll}{\mathcal{L}}
\newcommand{\Lc}{\mathsf{L}}
\newcommand{\Kc}{\mathcal{K}}

\newcommand{\Uc}{\mathsf{U}}

\newcommand{\Gc}{\mathsf{G}}

\newcommand{\lb}{\boldsymbol{\ell}}
\newcommand{\diag}{diag}
\DeclareMathOperator*{\esup}{ess\,sup}
\DeclareMathOperator*{\einf}{ess\,inf}
\newcommand{\mb}{\mathbf}
\newcommand{\xb}{\mathbf{x}}

\newcommand{\supp}{\operatorname{supp}}
\newcommand{\grad}{\operatorname{grad}}

\newcommand{\rango}{\operatorname{rank}}

\renewcommand{\diag}{\operatorname{diag}}
\title{{\bf On the existence of compactly supported reconstruction functions in a sampling problem}}
\author{{\bf A.~G. Garc\'{\i}a,}\thanks{E-mail:\texttt{agarcia@math.uc3m.es}} \,\,
{\bf M.~A. Hern\'andez-Medina,}\thanks{E-mail:\texttt{mahm@mat.upm.es}}\,\,
{\bf and  G.~P\'erez-Villal\'on}\thanks{E-mail:\texttt{gperez@euitt.upm.es}}}
\date{}
\begin{document}
\maketitle
\begin{itemize}
\item[*] Departamento de Matem\'aticas, Universidad Carlos III de Madrid,  
Avda. de la Universidad 30, 28911 Legan\'es-Madrid, Spain.
\item[\dag] Departamento de Matem\'atica Aplicada, E.T.S.I.T., U.P.M.,    
 Avda. Complutense s/n , 28040 Madrid, Spain.
 \item[\ddag] Departamento de Matem\'atica Aplicada, E.U.I.T.T., U.P.M.,    
 Carret. Valencia km.~7, 28031 Madrid, Spain.
\end{itemize}
\begin{abstract}
Assume that  samples of a filtered version of a function in a shift-invariant space are avalaible. This work deals with the existence of a sampling formula involving these samples and having reconstruction functions with compact support. Thus, low computational complexity is involved and 
truncation errors are avoided.  This is done in the light of the generalized sampling theory by using the oversampling technique: more samples than strictly necessary are used.  For a suitable choice of the sampling period, a necessary and sufficient condition is given in terms of the Kronecker canonical form of a matrix pencil. Comparing with other  characterizations in the mathematical literature, the given here has an important advantage: it can be reliable computed by using the GUPTRI form of the matrix pencil. 
Finally, a practical method for computing the compactly supported reconstruction functions is given for the important case where the oversampling rate is minimum.
\end{abstract}

\noindent{\bf AMS}: 15A21; 15A22; 42C15; 42C40; 94A20.

\section{Statement of the problem}

Let $V_\varphi$ be a shift-invariant space in $L^2(\RR)$ with stable generator $\varphi \in L^2(\RR)$, i.e.,
\[
V_\varphi:=\Big\{f(t)=\sum_{n\in \ZZ} a_n \ \varphi(t -n)\,:\, \{a_n\}\in
\ell^2(\ZZ)\Big\}\subset L^2(\RR)\,.
\]
Nowadays, sampling theory in shift-invariant spaces is a very active research topic (see, for instance, 
\cite{aldroubi:02,aldroubi:01,aldroubi:05,aldroubi:94,aldroubi:92,djo:97,walter:92,sun:99} and the references therein) since an appropriate choice for the generator $\varphi$ (for instance, a B-spline)
eliminates most of the problems  associated with the classical  Shannon's sampling theory 
\cite{unser:99}.

Suppose that a linear time-invariant system $\mathcal{L}$ is defined on $V_\varphi$. Under suitable conditions, Unser and Aldroubi \cite{aldroubi:94,unser:94} have found sampling formulas allowing  the recovering of any function $f\in V_\varphi$ from the sequence of samples 
$\{\big(\mathcal{L}f \big) (n)\}_{n\in \ZZ}$. Concretely, they proved that  for any $f\in V_\varphi$,
\begin{equation}
\label{samp}
f(t)=\sum_{n\in\ZZ}\mathcal{L}f(n)S_{\mathcal{L}}(t-n),\quad t\in\RR\,,
\end{equation}
where the sequence $\{S_{\mathcal{L}}(t-n)\}_{n\in \ZZ}$ is a Riesz basis for $V_\varphi$. Note that a reconstruction function $S_{\mathcal{L}}$  with compact support  implies low computational complexity and avoids truncation errors. Even when the generator $\varphi$ has compact support, rarely the same occurs with the reconstruction function $S_{\mathcal{L}}$ in formula \eqref{samp}. 
A way to overcome this difficulty is to use the oversampling technique, i.e, to take samples 
$\{\big(\mathcal{L}f\big)(nT)\}_{n\in \ZZ}$ with  a sampling period $T<1$. This is the main goal in this  paper: Assuming that both the generator $\varphi$ and $\mathcal{L} \varphi$ have compact support, we study the existence of stable sampling formulas  with compactly supported reconstruction functions, which allow us to recover any $f\in V_\varphi$ from the samples 
$\{\big(\mathcal{L}f\big)(nT)\}_{n\in \ZZ}$, where the sampling period is $T:=r/s<1$ for some positive integers $r<s$. This is done in the light of the generalized sampling theory obtained in \cite{gerardo:05b} by following an idea of Djokovic and Vaidyanathan in \cite{djo:97}.
For the sake of notational simplicity we have assumed that only samples from one linear time-invariant system $\mathcal{L}$ are avalaible.

In so doing, the problem is connected with the search of polynomial left inverses of a certain 
$s\times r$ polynomial matrix $\mathsf{G}(z)$ intimately related to the sampling problem. Taking advantage of the special structure of the matrix $\mathsf{G}(z)$ we give a necessary and sufficient condition which involves the matrix pencils theory. Concretely, this condition uses some information contained in  the Kronecker canonical form of a matrix pencil associated with the matrix $\mathsf{G}(z)$. From a practical point of view, this information can be retrieved from the GUPTRI (General UPper TRIangular) form of the matrix pencil. It is worth to mention that the GUPTRI form can be stably computed.

The mathematical problem of finding a polynomial left inverse of a polynomial matrix $\mathsf{G}(z)$ 
has been studied in \cite{cvetkovic:98} by Cvetkovi\'c and Vetterli  in the filter banks setting. It involves the Smith canonical form $\mathsf{S}(z)$ of the matrix $\mathsf{G}(z)$. Roughly speaking, the Smith canonical form of $\mathsf{G}(z)$ must contain monomials in its diagonal. From a practical point of view, the Smith canonical form has an important drawback: there is not a stable method for its computation.

Another algebraic approach is the following (see, for instance, \cite{raj:03}): Assume that $\mathsf{G}(z)$ is a $s\times r$ Laurent polynomial matrix ($r<s$); whenever the greatest common divisor of all minors of maximum order $r$ is a monomial, then its Smith canonical form $\mathsf{S}(z)$ has monomials in its diagonal. Without loss of generality we can assume that the $\gamma:=\binom{s}{r}$ minors of order $r$ in $\mathsf{G}(z)$ are polynomials with positive powers in $z$. Invoking Euclides algorithm we can obtain $\binom{s}{r}$ polynomials, 
$f_1(z),\dots, f_\gamma(z)$, such that
\[
\sum_{n=1}^\gamma f_n(z) A_n(z)=m(z)\,,\quad \text{for all $z\in\CC$}
\]
where $A_n$, $1\leq n\leq \gamma$, are the minors of order $r$ of $\mathsf{G}(z)$ and $m(z)$ is a monomial. Denote by $D_n'(z)$ the adjoint matrix corresponding to the minor $A_n$ and $D_n(z)$
the matrix obtained from $D'_n(z)$ by adding $s-r$ zero columns. Thus, 
$D_n(z)\mathsf{G}(z)=A_n(z) \mathbf{I}_r$, and consequently
\[
\bigg(\sum_{n=1}^\gamma f'_n(z) D_n(z)\bigg)\mathsf{G}(z)=\mathbf{I}_r\,,
\]
where $f'_n(z):=f(z)/m(z)$ could be a Laurent polynomial, $1\leq n\leq \gamma$. From a practical point of view the drawback here is the effective calculation of the $\binom{s}{r}$ minors of $\mathsf{G}(z)$ whenever $r$ becomes larger.

The paper is organized as follows: In Section 2 we include the needed preliminaries to understand the raised problem. The existence of reconstruction functions with compact support depends on the rank, for $z\in \CC\setminus\{0\}$, of a polynomial matrix $\mathsf{G}(z)$, associated with the sampling problem. In Section 3, a suitable choice of the sampling period $T=r/s$ reduces our problem to a matrix pencil problem. Thus, we give a necessary and sufficient condition for the existence of compactly supported reconstruction functions which involves the Kronecker canonical form of a singular matrix pencil. Section 4 is devoted to compute a polynomial left inverse of the matrix $\mathsf{G}(z)$ in the important case where the oversampling rate is minimum. Finally, we briefly remind, as an Appendix,  the canonical forms alluded in what follows.

\section{Preliminaries on generalized sampling}

{}From now on, the function $\varphi \in L^2(\RR)$ is a stable generator for the shift-invariant space
\[
V_\varphi:=\Big\{f(t)=\sum_{n\in \ZZ} a_n \ \varphi(t -n)\,:\, \{a_n\}\in
\ell^2(\ZZ)\Big\}\subset L^2(\RR)\,,
\]
i.e., the sequence $\{\varphi(\cdot  -n)\}_{n \in \ZZ}$ is a Riesz basis 
for $V_\varphi$. A Riesz basis in a separable Hilbert space is the image of 
an orthonormal basis by means of a bounded invertible operator. 
Recall that the sequence $\{\varphi(\cdot  -n)\}_{n \in \ZZ}$ 
is a Riesz basis for $V_\varphi$ if and only if
\[
0 < \|\Phi\|_0 \le \|\Phi\|_\infty < \infty \,,
\]
where $\|\Phi\|_0$ denotes the essential infimum of the function $\Phi(w):=\sum_{k\in \ZZ} 
|\widehat{\varphi}(w+k)|^2$ in $(0,1)$, and $\|\Phi\|_\infty$ its essential 
supremum ($\widehat{\varphi}$ denotes, as usual, the Fourier transform of $\varphi$).
Furthermore, $\|\Phi\|_0$ and $\|\Phi\|_\infty$ are the optimal Riesz 
bounds \cite[p. 143]{ole:03}.

We assume throughout the paper that the functions in the shift-invariant space 
$V_\varphi$ are continuous on $\RR$. Equivalently, the 
generator $\varphi$ is continuous on $\RR$ and the function $\sum_{n\in \ZZ} |\varphi(t-n)|^2$ 
is uniformly 
bounded on $\RR$ (see \cite{sun:99}). Thus, any $f\in V_\varphi$ is defined  as the 
pointwise sum $f(t)=\sum_{n\in \ZZ} a_n \varphi(t-n)$ on $\RR$. Besides, $V_\varphi$ is a reproducing kernel Hilbert space where convergence in the $L^2(\RR)$-norm implies pointwise convergence which is uniform on $\RR$ (see \cite{gerardo:05b}).

The space $V_\varphi$ is the image of $L^2(0,1)$ by means of the
isomorphism $\mathcal{T}_\varphi : L^2(0,1)\rightarrow V_\varphi$ which maps the
orthonormal basis $\{e^{-2\pi inw}\}_{n\in \ZZ}$ for $L^2(0,1)$ onto the
Riesz basis $\{\varphi(t-n)\}_{n\in\ZZ}$ for $V_\varphi$. Namely, for each $F \in L^2(0,1)$ the function $\mathcal{T}_\varphi F \in V_\varphi$ is given by
\begin{equation}
\label{iso}
(\mathcal{T}_\varphi F)(t):=\sum_{n\in\ZZ} \big\langle F(\cdot),e^{-2\pi in\cdot}\big \rangle_{L^2(0,1)}
\varphi(t-n),\quad t\in \RR\,.
\end{equation}
Suppose that $\mathcal{L}$ is a linear time-invariant system defined on $V_\varphi$ of  one of the following types (or a linear combination of both):
\begin{enumerate}[(a)]
\item The impulse response $\mathsf{h}$ of $\mathcal{L}$ belongs to $L^1(\RR)\cap L^2(\RR)$. 
Thus, for any $f \in V_\varphi$ we have
\[
\big(\mathcal{L} f\big)(t):=[f\ast \mathsf{h}](t)=\int_{-\infty}^\infty f(x)\mathsf{h}(t-x)dx\,,
\quad t \in \mathbb{R}\,.
\]
\item $\mathcal{L}$ involves samples of the function itself, i.e., $(\mathcal{L} f)(t)=f(t+d)$, $ t\in\RR$,
for some constant $d\in\RR$.
\end{enumerate}

For  fixed positive integers $s>r$, consider the sampling period $T:=r/s <1$.
The first goal is to recover any function $f\in V_\varphi$  by using a frame expansion involving the samples $\big\{(\mathcal{L} f)(rn/s)\big\}_{n\in \ZZ }$. This can be done in the light of the generalized sampling theory developed in \cite{gerardo:05b}. Indeed, since the sampling points $r n/s$, $n\in \ZZ$, can be expressed as
\[
\big\{r n/s\big\}_{n\in\ZZ}=
\big\{rm+(j-1)r/s \big\}_{m\in\ZZ,j=1,2,\ldots,s}\,,
\]
the initial problem is equivalent to the recovery of $f\in V_\varphi$ from the sequences of samples
\[
\{\mathcal{L}_jf\big(r n\big)\}_{n\in\ZZ,j=1,2,\ldots,s}
\]
where the linear time-invariant systems $\mathcal{L}_j$, $j=1,2,\ldots,s$, are defined by
\[
(\mathcal{L}_j f)(t):=(\mathcal{L}f)\big[t+(j-1)r/s\big]\,,\quad t\in \RR\,.
\]
Following the notation introduced in \cite{gerardo:05b}, consider the functions $g_j\in L^2(0,1)$, 
$j=1,2,\ldots,s$, defined as
\begin{equation}
\label{defgj}
g_j(w):= \sum_{n\in\ZZ} (\mathcal{L}\varphi)\big[ n+(j-1)r/s \big] e^{-2\pi inw} \,,
\end{equation}
the $s\times r$ matrix
\[
\mathbf{G}(w):=
\begin{bmatrix} g_1(w)& g_1(w+\frac{1}{r})&\cdots&g_1(w+\frac{r-1}{r})\\
g_2(w)& g_2(w+\frac{1}{r})&\cdots&g_2(w+\frac{r-1}{r})\\
\vdots&\vdots&&\vdots\\
g_s(w)& g_s(w+\frac{1}{r})&\cdots&g_s(w+\frac{r-1}{r})
\end{bmatrix}=
\bigg[g_j\Big(w+\frac{k-1}{r}\Big)\bigg]_{\substack{j=1,2,\ldots,s \\
k=1,2,\ldots, r}}\,,
\]
and its related constants
\[
\alpha_{\mathbf{G}}:=\einf_{w \in (0,1/r)}\lambda_{\min}[\mathbf{G}^*(w)\mathbf{G}(w)],\quad
\beta_{\mathbf{G}}:=\esup_{w \in (0,1/r)}\lambda_{\max}[\mathbf{G}^*(w)\mathbf{G}(w)]\,,
\]
where $\mathbf{G}^*(w)$ denotes the transpose conjugate of the matrix 
$\mathbf{G}(w)$, and $\lambda_{\min}$ (respectively $\lambda_{\max}$) the 
smallest (respectively the largest) eigenvalue of the positive semidefinite matrix 
$\mathbf{G}^*(w)\mathbf{G}(w)$. 
Notice that in the definition of the matrix $\mathbf{G}(w)$ we are considering the $1$-periodic 
extensions of the involved functions $g_j$, \ $j=1,2,\ldots,s$. 

Thus, the generalized sampling theory in \cite{gerardo:05b} (see Theorem 1, Theorem 2 and 
its proof)  gives the following sampling result in $V_\varphi$:

\begin{theorem}
\label{sampling1} 
Assume that the functions $g_j $, $j=1,2,\ldots,s$, defined in \eqref{defgj} belong to $L^\infty(0,1)$ (this is equivalent to $\beta_{\mathbf{G}}<\infty$). 
Then the following statements are equivalent:
\begin{enumerate}[(i)]
\item  $\alpha_{\mathbf{G}}>0$
\item There exist functions $a_j$ in $L^\infty(0,1)$, $j=1,2,\ldots,s$, such that
\begin{equation}
\label{a} 
\big[a_1(w),\ldots,a_s(w)\big]\, \mathbf{G}(w)=[1,0,\ldots,0]\,\,  \text{ a.e. in } (0,1)\,.
\end{equation}
\item There exists a frame for $V_\varphi$ having the form $\{S_j(\cdot-rn)\}_{n\in\ZZ,j=1,2,\ldots,s}$ such that, for any  $f\in V_\varphi$, we have
\begin{equation}
\label{s1} 
f=\sum_{n\in\ZZ}\sum_{j=1}^s(\mathcal{L}f)\big[rn+(j-1)r/s\big]\ S_j\big(\cdot-rn\big) \,\, 
\text{ in } L^2(\RR)\,.
\end{equation}
\end{enumerate}
In case the equivalent conditions are satisfied, the reconstruction functions in \eqref{s1} are given by 
$S_j=r\mathcal{T}_\varphi a_j$, $j=1,2,\ldots,s$, where the functions $a_j$, $j=1,2,\ldots,s$, 
satisfy \eqref{a}.
The convergence of the series in \eqref{s1} is also absolute and uniform on $\RR$.
\end{theorem}

Recall that a sequence $\{f_k\}$ is a frame for a separable Hilbert space $\mathcal{H}$ if there exist two constants $A,B>0$ (frame bounds) such that
\[
A\|f\|^2 \leq \sum_k |\langle f, f_k \rangle|^2 \leq B \|f\|^2 \,\, \text{ 
for all } f\in \mathcal{H} \,.
\]
Given a frame $\{f_k\}$ for $\mathcal{H}$ the representation 
property of any vector $f\in \mathcal{H}$ as a series $f=\sum_k c_k f_k$ is retained, 
but, unlike the case of Riesz bases, the uniqueness 
of this representation (for overcomplete frames) is sacrificed. 
Suitable frame coefficients 
$c_k$ which depend continuously and linearly on $f$ 
are obtained by using the dual frames $\{g_k\}$ of $\{f_k\}$, i.e., 
$\{g_k\}$ is another frame for $\mathcal{H}$ such that 
$f=\sum_k \langle f, g_k \rangle f_k=\sum_k \langle f, f_k \rangle g_k $ for each $f\in \mathcal{H}$. For
more details on the frame theory see the superb monograph \cite{ole:03} and the references therein.

\medskip

It is worth to mention that whenever  the $1$-periodic functions $g_j$, $j=1,2,\ldots,s$, are continuous on 
$\RR$,  the conditions in Theorem \ref{sampling1} are also equivalent to the condition recently introduced in \cite[Corollary 1]{gerardo:08a}:
\[
(iv)\qquad \rango \mathbf{G}(w)=r\,\, \text{ for all } w\in\RR\,.
\]

\medskip

The goal in this paper is to obtain necessary and sufficient conditions assuring that we can find reconstruction functions $S_j$, $j=1,2,\ldots,s$, in formula 
\eqref{s1} having compact support. To this end,  assume from now on that the generator $\varphi$ and 
$\mathcal{L}\varphi$ are compactly supported. 
We introduce the $s\times r$ matrix
\begin{equation}
\label{matrixG}
\mathsf{G}(z):=
\begin{bmatrix} \mathsf{g}_1(z)& \mathsf{g}_1(Wz)&\cdots&\mathsf{g}_1(W^{r-1}z)\\
 \mathsf{g}_2(z)& \mathsf{g}_2(Wz)&\cdots&\mathsf{g}_2(W^{r-1}z)\\
\vdots&\vdots&&\vdots\\
 \mathsf{g}_s(z)& \mathsf{g}_s(Wz)&\cdots&\mathsf{g}_s(W^{r-1}z)
\end{bmatrix}
\end{equation}
where $\displaystyle \,W:=e^{-2\pi i/r}\,$ and
$\mathsf{g}_j(z):= \sum_{n\in\ZZ} (\mathcal{L}\varphi)\big[n+(j-1)r/s\big]z^{n}$, $j=1,2\ldots,s$.
Note that  the matrix $\mathsf{G}(z)$ has Laurent polynomials entries, and 
$\mathbf{G}(w)=\mathsf{G}(e^{-2\pi iw})$. On the other hand, if the functions
$\mathsf{a}_j(z)$, $j=1,2\ldots,s$,  are Laurent polynomials satisfying
\begin{equation}
\label{aa}
[\mathsf{a}_1(z),\ldots,\mathsf{a}_s(z)]\mathsf{G}(z)=[1,0,\ldots,0]\,,
\end{equation}
then, the trigonometric polynomials $a_j(w)=\mathsf{a}_j(e^{-2\pi iw})$, 
$j=1,2,\ldots,s$,  satisfy \eqref{a}. In this case, the corresponding reconstruction functions
$S_j$, $j=1,2,\ldots,s$, have compact support. Indeed, in terms of the coefficients $c_{j,n}$ of 
$\mathsf{a}_j(z)$, that is, $\mathsf{a}_j(z)=\sum_{n\in\ZZ}c_{j,n}z^n$, $j=1,2,\ldots,s$, the reconstruction function $S_j$, $j=1,2,\ldots,s$,  can be written as (see \eqref{iso}):
\begin{equation}
\label{S_j}
S_j(t)=r \sum_{n\in \ZZ}c_{j,n}\varphi(t-n)\,,\quad t\in \RR\,.
\end{equation}

The existence of polynomial solutions of \eqref{aa} is equivalent to the existence of a left inverse of the matrix $\mathsf{G}(z)$ whose entries are polynomials. This problem has been studied in 
\cite{cvetkovic:98} by Cvetkovi\'c and Vetterli  in the filter banks setting. By using the Smith canonical form $\mathsf{S}(z)$ of the matrix $\mathsf{G}(z)$ (see Appendix \ref{smith}), a characterization for the  existence of polynomial solutions of \eqref{aa}  has been found in \cite{garcia:08}.  Namely,  assuming  that the generator 
$\varphi$ and $\mathcal{L}\varphi$ have compact support, there exists a
polynomial vector  $[\mathsf{a}_1(z),\mathsf{a}_2(z),\cdots,\mathsf{a}_s(z)]$ satisfying
\eqref{aa} if and only if the polynomials $i_j(z)$, $j=1,2,\ldots,r$, on the diagonal of the Smith canonical form $\mathsf{S}(z)$  of the matrix $\mathsf{G}(z)$ are monomials. Assume that the $s\times r$ matrix
\begin{equation}
\label{smithG}
\mathsf{S}(z)=\begin{bmatrix} i_1(z)& 0&\cdots&0\\
 0& i_2(z)&\cdots&0\\
\vdots&\vdots&&\vdots\\
0& 0&\cdots&i_r(z)\\
0& 0& \cdots & 0\\
\vdots&\vdots&&\vdots\\
0& 0&\cdots&0
\end{bmatrix}
\end{equation}
is the Smith canonical form of the matrix $\mathsf{G}(z)$ (note that it is the case whenever 
$\alpha_{\mathbf{G}}>0$) and consider the unimodular matrices 
$\mathbf{V}(z)$  and $\mathbf{W}(z)$, of dimension $s\times s$ and $r\times r$ respectively,  such that 
$\mathsf{G}(z)=\mathbf{V}(z)\mathsf{S}(z)\mathbf{W}(z)$. 

Observe that if $\mathsf{S}(z)$ is the Smith form of the matrix $\mathsf{G}(z)$ then, taking into account that $\mathbf{V}(z)$ and $\mathbf{W}(z)$ are unimodular matrices, we have
\[
\rango\mathsf{S}(z)=\rango \mathsf{G}(z) \,\, \text{for all}\,\, z\in \CC.
\]
Therefore, it is straightforward to deduce that, for each $j=1,2,\ldots,r$, the polynomial $i_j(z)$  is a monomial if and only if $\rango \mathsf{S}(z)=r$ for all $z\in\CC\setminus\{0\}$. This condition, under the above hypotheses on $\varphi$ and $\mathcal{L}\varphi$,  is equivalent to saying that 
\begin{equation}
\label{condicionrango}
\rango \mathsf{G}(z)=r \,\, \text{for all}\,\, z\in \CC\setminus\{0\}\,.
\end{equation}
(See \cite{garcia:08} for the details). The main aim in this work is to search for an equivalent condition to \eqref{condicionrango} useful from a practical point of view.

\section{Searching for an useful equivalent condition}

The first step is to reduce our polynomial matrix $ \mathsf{G}(z)$ to a matrix pencil in order to use the well-established theory.  In so doing we need some preliminaries.
Let $f(z)=a_m z^m+a_{m-1}z^{m-1}+\cdots+a_1z^1+a_0$ be an algebraic polynomial of order $m$,  and let $n$ be a positive integer. For each $j=0,1,\dots,n-1$ let $\widehat{f}_j(z)$ denote the sum of the monomials $a_rz^r$ where 
$r\equiv j(\mod n)$. Obviously, $f(z)=\sum_{j=0}^{n-1} \widehat{f}_j(z)$. The polynomial $\widehat{f}_j$, $0\leq j \leq n-1$, is the so-called $n$-harmonic of order $j$ of the polynomial $f$; it satisfies:
\[
\widehat{f}_j(e^{2\pi i /n}z)=e^{2\pi i j/n}\widehat{f}_j(z)\,\, \text{ for all  $z\in\CC$.}
\]

\medskip
 
Assume that $\supp \mathcal{L}\varphi$ is contained in an interval $[0,N]$, where $N\in \NN$.  Thus, the functions $\mathsf{g}_j(z)$  are Laurent polynomials in the variable $z$. Consider
\[
p:=\min\{q\in \NN : q\, \frac{r}{s}>1\}\,.
\]
It is easy to check that $p=c+1$ where $c$ denotes the quotient in the euclidean division $s|r$. Hence, we can write the Laurent polynomials $\mathsf{g}_i(z)$, $j=1,2 \ldots,s$, as:
\begin{align}
\label{sistema_g}
&\mathsf{g}_1(z)= \mathcal{L}\varphi(1)z+\mathcal{L}\varphi(2)z^2+\cdots+\mathcal{L}\varphi(N-1)z^{N-1}\notag\\
&\mathsf{g}_2(z)= \mathcal{L}\varphi \big(\frac{r}{s}\big)+\mathcal{L}\varphi\big(1+\frac{r}{s}\big)z+\cdots+\mathcal{L}\varphi\big(N-1+\frac{r}{s}\big)z^{N-1}\notag \\
 & \vdots                                                                             \notag    \\
&\mathsf{g}_p(z)= \mathcal{L}\varphi\big((p-1)\frac{r}{s}\big)+\mathcal{L}\varphi\big(1+(p-1)\frac{r}{s}\big)z+
\cdots+\mathcal{L}\varphi\big(N-1+(p-1)\frac{r}{s}\big)z^{N-1}
\\
&\mathsf{g}_{p+1}(z)=
\mathcal{L}\varphi\big(p \frac{r}{s}-1\big)z^{-1}+\cdots+\mathcal{L}\varphi\big(N-2+p \frac{r}{s}\big)z^{N-2}\notag
\\
&\vdots                                                                                  \notag    \\
&\mathsf{g}_s(z)= \mathcal{L}\varphi\big((s-1)\frac{r}{s}-r+1\big)z^{-(r-1)}+\cdots+\mathcal{L}\varphi\big(N-r+2+(s-1)\frac{r}{s}\big)z^{N-r+2}\,. \notag 
\end{align}
The polynomial $\mathsf{g}_1(z)$ has at most $N-1$ nonzero terms; the rest of polynomials $\mathsf{g}_j(z)$, $2\leq j\leq s$, have at most $N$  nonzero terms. In what follows,  we use the new matrix 
$\GG(z)=\mathsf{G}(z)\mathsf{U}(z)$, where 
\[
\mathsf{U}(z)=\diag\big[z^{(r-1)},(Wz)^{(r-1)},(W^2z)^{(r-1)},\dots,(W^{r-1} z)^{(r-1)}\big]\,.
\]
Thus, all the entries of the polynomial matrix $\GG(z)$ are algebraic polynomials in $z$ and,  moreover we have
$\rango\GG(z)=\rango\Gc(z)$ for all $z\in\CC\setminus\{0\}$. We denote by $\widetilde{\mathsf{g}}_j(z)$ the algebraic  polynomial $z^{r-1}\mathsf{g}_j(z)$, $1\leq j\leq s$.

\medskip

The strategy is to reduce the polynomial matrix $\GG(z)$ into another simpler one having the same rank  for all $z\in\CC\setminus\{0\}$. 

\begin{lemma}
\label{G->armonicos}
Consider the matrix $\widehat{\GG}(z)=[\widehat{\GG}_0(z)\, \widehat{\GG}_2(z)\, \dots \widehat{\GG}_{(r-1)}(z)]$, where 
$\widehat{\GG}_j(z)$, $0\leq j\leq (r-1)$, denotes the column vector consisting of the $r$-harmonics of order $j$ of the polynomials $\widetilde{\mathsf{g}}_i(z)$ where $1\leq i\leq s$. Then
\[
 \GG(z)=\widehat{\GG}(z)\Omega_r
\]
where $\Omega_r$ denotes the Fourier matrix  of order $r$.
\end{lemma}
\noindent{\bf Proof:} For each $i=1,2,\dots,s$ we have that
$\widetilde{\mathsf{g}}_i(z)=\sum_{j=0}^{r-1} \widehat{\widetilde{\mathsf{g}}}_{ij}(z)$ where $\widehat{\tilde{\mathsf{g}}}_{ij}(z)$ denotes the $r$-harmonic of order $j$ of $\widetilde{\mathsf{g}}_i$. We can write the matrix $\GG(z)$ as 
\begin{multline*}
   \GG(z)=\big[\widehat{\GG}_0(z)+\widehat{\GG}_1(z)+\cdots+\widehat{\GG}_{r-1}(z) \\
           \widehat{\GG}_0(z)+W\widehat{\GG}_1(z)+\cdots+W^{r-1}\widehat{\GG}_{r-1}(z) \\
           \cdots   \cdots  \\
           \widehat{\GG}_0(z)+W^{r-1}\widehat{\GG}_1(z)+\cdots+W^{(r-1)^2}\widehat{\GG}_{r-1}(z)\big]
\end{multline*}
Hence, in matrix form we have
\[
\GG(z)=\big[\widehat{\GG}_0(z)\,\, \widehat{\GG}_1(z)\,\,\dots \widehat{\GG}_{r-1}(z)\big]\Omega_r=
       \widehat{\GG}(z)\Omega_r	
\]
where
\[
\Omega_r=\begin{bmatrix}
	   1 & 1 & 1 & \cdots & 1\\
           1 & W & W^2& \cdots & W^{r-1}\\
	   1 & W^2 & W^4& \cdots & W^{2(r-1)}\\
           \vdots & & & \vdots \\
           1 & W^{r-1}& W^{2(r-1)} & \cdots & W^{(r-1)^2}
         \end{bmatrix}
\]
is the Fourier matrix of order $r$.
\mbox{}\hfill$\square$\medskip

\noindent Observe that $\rango\GG(z)=\rango\widehat{\GG}(z)$ for all $z\in\CC\setminus\{0\}$.

\medskip

In what follows, we assume that $\supp \mathcal{L}\varphi \subseteq [0,N]$ and, in addition, we also assume that
$N\leq r$. In this case, having in mind the number of nonzero consecutive terms of the polynomial 
$\widetilde{\mathsf{g}}_j(z)$,
we conclude that the $r$-harmonic of order $q$, $q=0,1\dots,r-1$, of the polynomial $\widetilde{\mathsf{g}}_i(z)$, 
$1\leq i\leq s$,  is a monomial having the form $c_{ip}z^{kr+q}$ where $c_{iq}\in\CC$ and $k\in\{0,1\}$. This choice of
$r$ and,  consequently, of the sampling periods $T=r/s$, $r,s\in\NN$ and $s > r$, simplifies the structure of the matrix
$\widehat{\GG}(z)$.

\medskip

First, let us to give an illustrative example:  Consider  $N=3$, $r=4$ and $s=5$; here $T=4/5$, $p=2$ and the polynomials 
$\widetilde{\mathsf{g}}_j(z)$ read
\begin{align}
\tilde{\mathsf{g}}_1(z)&= \star z^4+\star z^5 & \tilde{\mathsf{g}}_2(z)&= \star z^3+\star z^4+\star z^5 \notag\\
\tilde{\mathsf{g}}_3(z)&= \star z^2+\star z^3+\star z^4 & \tilde{\mathsf{g}}_4(z)&= \star z+\star z^2+\star z^3 \notag \\
\tilde{\mathsf{g}}_5(z)&= \star +\star z +\star z^2 & & \notag
\end{align} 
Hence, the matrix $\widehat{\GG}(z)$ reads
\begin{equation}
 \label{example2}
\widehat{\GG}(z)=
\begin{bmatrix}
  \star z^4 & \star z^5 & 0 &  0\\
  \star z^4 & \star z^5 & 0 & \star z^3\\
   \star z^4 & 0 & \star z^2 & \star z^3\\
   0 & \star z & \star z^2 & \star z^3\\
   \star & \star z & \star z^2 &  0\\
\end{bmatrix}
\end{equation}
This example shows that the 3rd and 4th columns have the form $z^2C$ and $z^3C'$ where $C,C'\in\CC^{s\times 1}$. The first and second columns do not share this property. If we right multiply the matrix
$\widehat{\GG}(z)$ by $\diag[1, z^{-1},z^{-2},z^{-3}]$,  we get the new matrix
\[
 \widetilde{\GG}(z):=
\begin{bmatrix}
  \star z^4 & \star z^5 & 0 &  0\\
  \star z^4 & \star z^5 & 0 & \star z^3\\
   \star z^4 & 0 & \star z^2 & \star z^3\\
   0 & \star z & \star z^2 & \star z^3\\
   \star & \star z & \star z^2 &  0\\
\end{bmatrix}
\begin{bmatrix}
 1 & & & \\
  & z^{-1} &  &   \\
  &        & z^{-2} & \\
  &        &        & z^{-3}       
\end{bmatrix}=
\begin{bmatrix}
  \star z^4 & \star z^4 & 0 &  0\\
  \star z^4 & \star z^4 & 0 & \star \\
   \star z^4 & 0 & \star  & \star \\
   0 & \star  & \star  & \star \\
   \star & \star  & \star  &  0\\
\end{bmatrix}
\]

\medskip

Now we can go into the general case for the matrix $\widehat{\GG}(z)$. Having in mind  
Eqs.\eqref{sistema_g} and that $\widetilde{\mathsf{g}}_i(z)=z^{r-1}\mathsf{g}_i(z)$ we obtain:
\[
\max\big\{\grad\widetilde{\mathsf{g}}_j: 1\leq j\leq s\big\}=(N-1)+(r-1)=N+r-2<2r\,.
\]
Hence, the matrix $\widehat{\GG}(z)$ has the form
\[
\widehat{\GG}(z)=
\begin{bmatrix}
  c_{11}z^{k_{11}r} & c_{12}z^{k_{12}r+1} & \cdots & c_{1r}z^{k_{1r}r+(r-1)}\\
     \vdots             &     \vdots              &  \vdots & \vdots                              \\
  c_{s1}z^{k_{s1}r} & c_{s2}z^{k_{s2}r+1} & \cdots & c_{sr}z^{k_{sr}r+(r-1)}
\end{bmatrix}
\]
where the coefficients $k_{ij}\in\{0,1\}$.  We can easily obtain the following result:
\begin{lemma}
\label{formageneral}
Assume that $N>1$. Then, for each $1 \leq j \leq N-1$ there exist indices $i'\neq i$, $1\leq i,\, i'\leq s$, such that 
$k_{ij}\neq k_{i'j}$. Otherwise, for each $N \leq j \leq r$ it holds that $k_{ij}=k_{i'j}$ for all $1\leq i,\, i'\leq s$.
\end{lemma}

\medskip

Assume that $N>1$ and recall that $N\leq r$. The entries of the $j$th column  of the matrix 
$\widehat{\GG}(z)$, where $N \leq j \leq r$, have the form $\star z^{j-1}$ ($\star\in\CC$); they could have the form $\star z^{j-1}$ or $\star z^{r+(j-1)}$ whenever $1\leq j\leq N-1$. Dividing the $j$th column by 
$z^{j-1}$, obviously we obtain a matrix with the same rank than $\widehat{\GG}(z)$ for any 
$z\in\CC\setminus\{0\}$. Thus, we introduce the new polynomial matrix  $\widetilde{\GG}(z)$:
\[
\widetilde{\GG}(z):=\widehat{\GG}(z)Q(z)=\big[\MM(z)\,  \mathcal{G}\big]\,,
\]
where $\mathcal{G}\in\CC^{s\times (r-N+1)}$ denotes an scalar matrix and $Q(z):=\diag[1,z^{-1},\dots,z^{1-r}]$.
Whenever $\rango \mathcal G< r-N+1$, we have that $\rango\widetilde{\GG}(z)=\rango\widehat{\GG}(z)<r$ for all 
$z\in\CC\setminus\{0\}$ and, hence, there is no a polynomial left inverse for $\widehat{\GG}(z)$. In the case 
$\rango\mathcal{G}(z)=r-N+1$, there exists an invertible matrix $R\in\CC^{s\times s}$ such that
\[
    R\ \mathcal{G}= \begin{bmatrix}
                                         \mathcal{G}'\\
						0
                                     \end{bmatrix},
 \]
where $\mathcal{G}'\in \CC^{(r-N+1)\times (r-N+1)}$ is invertible. Thus,
\[
 R\ \widetilde{\GG}(z)=\big[R\MM(z) \, R\mathcal{G}\big]=\begin{bmatrix}
                                             						\MM_1(z) & \mathcal{G}'\\
 											\MM_2(z) &	0
                                            					   \end{bmatrix}
\]
The entries of the polynomial matrix $\MM(z)\in\CC^{s\times (N-1)}$ are of the form $\star z^r$ or constants;  denoting 
$\lambda=z^r$, the matrices $\MM_i(z)$, $i=1,2$,  can be expressed as
\[
  \MM_i(\lambda)=M_{i1}-\lambda M_{i2}
\]
donde $M_{1i}\in\CC^{(r-N+1)\times (N-1)}$ y $M_{2i} \in\CC^{(s-r+N-1)\times (N-1)}$. As a consequence, we have the following result:
\begin{lemma}
Assume that $\rango\mathcal{G}=r-N+1$. Then,  $\rango \mathsf{G}(z)=r$ for all $z\in\CC\setminus\{0\}$ if and only if 
$\rango \MM_2(\lambda)=N-1$ for all $\lambda\in\CC\setminus\{0\}$. 
\end{lemma}

\medskip

Next step is to characterize when the rank of the matrix $M_{21}-\lambda M_{22}$ equals $N-1$ for any 
$\lambda \in \CC\setminus\{0\}$. To this end, we use the  Kronecker canonical form (KCF herafter) of the matrix pencil $\MM_2(\lambda)$ (see the Appendix \ref{FCK}  for the details). By using the block structure notation $A\oplus B:=\diag(A,B)$, consider the KCF of the matrix pencil $\MM_2(\lambda)$, i.e.,
\[
\KK(\lambda):=S^{right}_{\MM_2}(\lambda)\oplus J_{\MM_2}(\lambda)\oplus N_{\MM_2}(\lambda)\oplus S^{left}_{\MM_2}(\lambda)
\]
where $S^{right}_{\MM_2}(\lambda)$ denotes the right singular part of $\MM_2(\lambda)$, 
$S^{left}_{\MM_2 }(\lambda)$ denotes the left singular part, $J_{\MM_2}(\lambda)$ is the block associated with the finite eigenvalues of the pencil and, finally, $N_{\MM_2}(\lambda)$ is the block associated with the infinite eigenvalue.
Having in mind the structure of the different blocks appearing in the KCF of the matrix pencil $\MM_2(\lambda)$, we can derive that the rank of $\KK(\lambda)$, and consequently of $\MM_2(\lambda)$, is $N-1$ for all $\lambda\in\CC\setminus\{0\}$ if and only if
$\KK(\lambda)$ has not right singular part and the only possibly finite eigenvalue is the zero one. In fact, we have the following result:
\begin{lemma}
\label{rangoHAZ}
The rank of matrix  $\MM_2(\lambda)$ is $N-1$ for each $\lambda\in\CC\setminus\{0\}$ if and only if the following conditions hold:
\begin{enumerate}
 \item The KCF of the matrix pencil $\MM_2(\lambda)$ has not right singular part and,
 \item If $\mu$ is a finite eigenvalue of the matrix pencil $\MM_2(\lambda)$, then $\mu=0$.
\end{enumerate}
\end{lemma}

\medskip

Now, Lemma \ref{rangoHAZ} allows us to decide when the rank of our initial polynomial matrix 
$\mathsf{G}(z)$ is $r$ for all 
$z\in\CC\setminus\{0\}$. Let us to remind all the given steps  in reducing the initial polynomial matrix 
$\mathsf{G}(z)$:
\[
\mathsf{G}(z)\rightsquigarrow \GG(z)\rightsquigarrow \widehat{\GG}(z)\rightsquigarrow \widetilde{\GG}(z)\rightsquigarrow \begin{bmatrix} \MM_1(z) & \mathcal{G}'\\
  								       \MM_2(z) &	0
                                            		\end{bmatrix}\,,
\]
where
\begin{enumerate}
 \item $\GG(z)=\mathsf{G}(z)\mathsf{U}(z)$,
 \item $\widehat{\GG}(z)\Omega_r=\GG(z)$, 
 \item $\widetilde{\GG}(z)=\widehat{\GG}(z)Q(z)=[\MM(z)|\mathcal{G}]$, where $\mathcal{G}\in\CC^{s\times (r-N+1)}$ and $Q(z)=\diag[1,z^{-1},\dots,z^{1-r}]$, 
 \item If $\rango\mathcal{G}=r-N+1$, there exists $R\in\CC^{s\times s}$ invertible such that
       $R\widetilde{\GG}(z)=\begin{bmatrix} \MM_1(z) & \mathcal{G}'\\ \MM_2(z) & 0\end{bmatrix}$ where the matrix
       $\mathcal{G}'\in\CC^{(r-N+1)\times (r-N+1)}$ is invertible,
 \item The matrices $\MM_i(z)$, $i=1,2$, can be expressed as $\MM_i(\lambda)=M_{i1}-\lambda M_{i2}$ with 
 $\lambda=z^r$.
\end{enumerate}
As a consequence, we have proved the following result:
\begin{theorem}
\label{principal}
Assume that $\supp \mathcal{L}\varphi \subseteq [0,N]$, where $N\in \NN$ with $N>1$, and take $N\leq r < s$. Let $\mathsf{G}(z)$ be the corresponding $s\times r$ polynomial matrix given in \eqref{matrixG}. Then, 
$\rango \mathsf{G}(z)= r$ for any $z \in \CC\setminus\{0\}$
if and only if the following statements hold:
\begin{enumerate}
 \item $\rango \mathcal{G}=r-N+1$ and,
 \item the KCF of the matrix pencil $\MM_2(\lambda)$ has not right singular part, and the only possible finite eigenvalue is $\mu=0$. 
\end{enumerate}
\end{theorem}

For practical purposes it is not necessary  to compute the KCF of the matrix pencil 
$\MM_2(\lambda)$ (if possible). The needed information about $\MM_2(\lambda)$ is obtained from its GUPTRI form (Generalized UPer TRIangular form). See the Appendix \ref{formaguptri} for the details. As the matrix 
$\widetilde{\GG}(z)$ depends on $z^r$, in what follows we identify the matrix
$\widetilde{\GG}(z)$ with $\widetilde{\GG}(\lambda)$ where $\lambda=z^r$.

\subsection{A toy model involving the quadratic B-spline}
\label{ejemploN3}
The following example illustrates the result given in Theorem  \ref{principal}. Consider as generator 
$\varphi$ the quadratic B-spline $N_3(x)$, i.e., 
\[
 N_3(x)=\frac{x^2}{2}\chi_{[0,1)}+(-\frac{3}{2}+3x-x^2)\chi_{[1,2)}+\frac{1}{3}(3-x)^2\chi_{[2,3)}\,,
\]
where $\chi_{[a,b)}$ denotes the characteristic function of the interval $[a,b)$. In this case, for the identity system $\mathcal{L}f=f$ for all $f\in V_\varphi$ we have $\supp\mathcal{L}\varphi\subseteq [0,3]$, i.e., $N=3$. Taking the sampling period $T=4/5$, i.e., $r=4$ and $s=5$, the Laurent polynomials $\mathsf{g}_i(z)$ given by \eqref{sistema_g} read:
\begin{align*}
{\mathsf{g}}_1(z)&= \frac{1}{2}z+\frac{1}{2}z^2 & {\mathsf{g}}_2(z)&= \frac{8}{25}+\frac{33}{50}z+\frac{1}{50}z^2 \\
{\mathsf{g}}_3(z)&= \frac{9}{50} z^{-1}+\frac{37}{50}+\frac{2}{25} z & {\mathsf{g}}_4(z)&= \frac{2}{25} z^{-2}+\frac{37}{50}z^{-1}+\frac{9}{50}  \\
{\mathsf{g}}_5(z)&=\frac{1}{50} z^{-3}+\frac{33}{50}z^{-2}+\frac{8}{25} z^{-1}& &  
\end{align*}
Following the above steps we obtain
\[
\widehat{\GG}(z)=\begin{bmatrix}
				\frac{1}{2}z^4 & \frac{1}{2}z^5 & 0 & 0\\[3pt]
				\frac{33}{50}z^4&\frac{1}{50}z^5 & 0 & \frac{8}{25}z^3\\[3pt]
				\frac{2}{25}z^4& 0				&\frac{9}{50}z^2 & \frac{37}{50}z^3\\[3pt]
				0			& \frac{2}{25}z& \frac{37}{50}z^2	& \frac{9}{50}z^3\\[3pt]
				\frac{1}{50}	& \frac{33}{50}z& \frac{8}{25}z^2	& 0
                   		\end{bmatrix}
\]
Right multiplication by the matrix $\diag[1,z^{-1},z^{-2},z^{-3}]$ gives:
\[
\widetilde{\GG}(\lambda)=[\MM(\lambda) \,|\, \mathcal{G}]=\left[\begin{array}{cc|cc}
				\frac{1}{2}\lambda & \frac{1}{2}\lambda & 0 & 0\\[3pt]
				\frac{33}{50}\lambda&\frac{1}{50}\lambda & 0 & \frac{8}{25}\\[3pt]
				\frac{2}{25}\lambda& 0			&\frac{9}{50} & \frac{37}{50}\\[3pt]
				0			& \frac{2}{25}& \frac{37}{50}	& \frac{9}{50}\\[3pt]
				\frac{1}{50}	& \frac{33}{50}& \frac{8}{25}	& 0
                   		\end{array}\right]
\]
where $\lambda=z^4$. The matrix $\mathcal{G}\in\CC^{5\times 2}$ has rank 2;  performing some elementary operations on the rows of 
$\mathcal{G}$ we obtain
\[
	\mathcal{G}'=\begin{bmatrix}
                             			 \frac{9}{50} & \frac{37}{50}\\[3pt]
						0			& \frac{8}{25}\\[3pt]
						0			& 0\\[3pt]
						0			& 0\\[3pt]
						0			& 0
                            			\end{bmatrix}=\begin{bmatrix}
			0 &\phantom{-} 0 & 1 &0 &0 \\[3pt]
			0 & \phantom{-}1 & 0 & 0&0 \\[3pt]
			1 & \phantom{-}0 & 0 & 0& 0\\[3pt]
			\frac{37}{9} & -\frac{161}{18} & 0 &1 &0 \\[3pt]
			\frac{16}{9} & -\frac{37}{9} & 0 &0 &1
	             \end{bmatrix}
				\begin{bmatrix}
				0 & 0\\[3pt]
				0 & \frac{8}{25}\\[3pt]
				\frac{9}{50} & \frac{37}{50}\\[3pt]
				\frac{37}{50}	& \frac{9}{50}\\[3pt]
				 \frac{8}{25}	& 0
                   		\end{bmatrix}=R\mathcal{G}
\]
Therefore, $R\widetilde{\GG}(\lambda)=[R\MM(\lambda)\, R\mathcal{G}]=\begin{bmatrix} \MM_1(\lambda) & \mathcal{G}'\\ \MM_2(\lambda) & 0 \end{bmatrix}$ where
\[
\MM_2(\lambda)=\begin{bmatrix}
		\phantom{\frac{1}{50}++}\frac{1}{2}\lambda & \phantom{\frac{1}{50}++}\frac{1}{2}\lambda\\[3pt]
		\phantom{\frac{1}{50}+}\frac{5017}{900}\lambda & \frac{2}{25}+\frac{161}{900}\lambda\\[3pt]
		\frac{1}{50}+\frac{1157}{450}\lambda & \frac{33}{50}+\frac{37}{450}\lambda
                 \end{bmatrix}
\]
In this case, a direct computation gives $\KK_{\MM_2}(\lambda)=L^\top_2(\lambda)$. As a consequence, Theorem \ref{principal} ensures that the corresponding polynomial matrix $\mathsf{G}(z)$ possess a polynomial left inverse.

Next we deal with the problem of computing a polynomial left inverse of $\mathsf{G}(z)$ whenever it does exist.

\section{Computing a  polynomial left inverse of the matrix  $\mathsf{G}(z)$}
First note that if we compute a polynomial left inverse of the matrix $\widetilde{\GG}(\lambda)$ then we obtain a polynomial left inverse of the matrix $\mathsf{G}(z)$. Indeed, remind that
\[
   \widetilde{\GG}(z)=\mathsf{G}(z)\mathsf{U}(z)\Omega_r^{-1}Q(z)
\]
where $\mathsf{U}(z)=\diag[z^{r-1},(Wz)^{r-1},\dots,(W^{r-1}z)^{r-1}]$, $\Omega_r$ is the Fourier matrix of order $r$, and
$Q(z)=\diag[1,z^{-1},\dots,z^{1-r}]$. Thus, if $L(z)$ is a polynomial left inverse of the matrix 
$\widetilde{\GG}(z)$, then the matrix
\[
   L_{\mathsf{G}}(z)=\diag[1,z,\dots,z^{r-1}]\Omega_r\diag[z^{1-r},(Wz)^{1-r},\dots,(W^{r-1}z)^{1-r}]L(z)
\]
will be a polynomial left inverse of the matrix $\mathsf{G}(z)$. Hence, we concentrate ourselves in computing a polynomial left inverse of the  matrix $\widetilde{\GG}(z)$. To this end, consider 
$\widetilde{\GG}(\lambda)=\Ac^\top - \lambda\Bc^\top$ 
($\lambda=z^r$); being $L(\lambda)$ a polynomial left inverse of the matrix $\widetilde{\GG}(\lambda)$, we have $(\Ac-\lambda\Bc)L^\top(\lambda)=\mathbf{I}_r$. Let us denote $\Lc(\lambda):=L^\top(\lambda)$. As we are searching for $s\times r$ matrices $\Lc(\lambda)$, whose entries are polynomials, such that 
$(\Ac-\lambda\Bc)\Lc(\lambda)=\mathbf{I}_r$ we can use the following notation:
\begin{align*}
 \Lc(\lambda)\phantom{\mbox{}^\top}&=\,\big[\Lc_1(\lambda)\, \Lc_2(\lambda)\, \dots \, \Lc_r(\lambda)\big]\,, \,\text{i.e.,\, $\Lc_i(\lambda)$ denotes the $i$th column of  $\Lc(\lambda)$},\\
 \Lc_i(\lambda)\phantom{\mbox{}^\top}&=\,\lb^0_i+\lb_i^1\lambda+\cdots+\lb_i^\nu\lambda^\nu,\quad i=1,2,\dots,r\,, \text{ where $\lb_i^k \in \CC^s$},\,\, k=0,1,\dots, \nu\,.
\end{align*} 
As a consequence, $(\Ac-\lambda\Bc)\Lc(\lambda)=\mathbf{I}_r$ is equivalent to
\begin{equation}
\label{Eq:v}
 \Ac\lb_i^0+(\Ac\lb_i^1-\Bc\lb_i^0)\lambda+ \cdots
      +(\Ac\lb_i^\nu-\Bc\lb_i^{\nu-1})\lambda^\nu-\Bc\lb_i^\nu\lambda^{\nu+1}=I_r^i\,,\quad i=1,2,\dots,r,
\end{equation}
where $I_r^i$ denotes  the $i$th column  of the identity matrix $\mathbf{I}_r$.  Equating coefficients, for each $i=1,2,\dots,r$, we obtain the set of linear  equations
\[
\Ac\lb_i^{0}=I_r^i\,,\,\,\,\Ac\lb_i^1-\Bc\lb_i^0=0\,,\ldots,\,\,\,\Ac\lb_i^\nu-\Bc\lb_i^{\nu-1}=0\,,\,\,\, -\Bc\lb_i^{\nu}=0\,,
\]
or in matrix form
\begin{equation}
\label{matAB2}
\begin{bmatrix}
 -\Bc &   & & & &  \\
 \Ac &-\Bc & & & &\\
     &\Ac & -\Bc & & &\\
     &    &     &\ddots & &\\
     &    &     &       &\Ac & -\Bc\\
     &    &     &       &    &\Ac
\end{bmatrix}
\begin{bmatrix}
 \lb_i^{\nu\phantom{-1}}\\
  \lb_i^{\nu-1}\\
    \vdots\\
 \lb_i^{0\phantom{-1}}\\
\end{bmatrix}=
\begin{bmatrix}
 \mathbf{0}\\
 \vdots \\
 \mathbf{0}\\
    \vdots\\
     \mathbf{0}\\
      I_r^i\\
\end{bmatrix},\quad i=1,2,\dots,r\,,
\end{equation}
where the resulting block matrix has order $(\nu+2)r\times (\nu+1)s$. The goal is to find $\nu\in \NN$ such that the above $r$ linear systems become consistent. Next, we come back to the example in Section \ref{ejemploN3}.

\noindent{\bf The example revisited:}  Consider again the example involving the quadratic B-spline given in Section \ref{ejemploN3}. In this case, 
$\widetilde{\GG}(z)=\Gc(z)\Uc(z)\Omega_4^{-1}\diag[1,z^{-1},z^{-2},z^{-3}]$ and, taking  $\lambda=z^4$ we have
\[
\widetilde{\GG}(\lambda)=\begin{bmatrix}
				\frac{1}{2}\lambda & \frac{1}{2}\lambda & 0 & 0\\[3pt]
				\frac{33}{50}\lambda&\frac{1}{50}\lambda & 0 & \frac{8}{25}\\[3pt]
				\frac{2}{25}\lambda& 0			&\frac{9}{50} & \frac{37}{50}\\[3pt]
				0			& \frac{2}{25}& \frac{37}{50}	& \frac{9}{50}\\[3pt]
				\frac{1}{50}	& \frac{33}{50}& \frac{8}{25}	& 0
                   		\end{bmatrix}=\Ac^\top-\lambda\Bc^\top\,,
\]
where
\[
\Ac=\begin{bmatrix}
     0 & 0 & 0 &0 & \frac{1}{50}\\[3pt]
     0 & 0 & 0 &\frac{2}{25} & \frac{33}{50}\\[3pt]
     0 & 0 & \frac{9}{50} &\frac{37}{50} & \frac{8}{50}\\[3pt]
     0 & \frac{8}{25} & \frac{37}{50} & \frac{9}{50}& 0
    \end{bmatrix}\quad \text{y}\quad
\Bc=\begin{bmatrix}
     -\frac{1}{2} & -\frac{33}{50} & -\frac{2}{25} &0 & 0\\[3pt]
     -\frac{1}{2} & -\frac{1}{50} & 0 & 0 & 0\\[3pt]
     0 & 0 & 0 & 0 & 0\\[3pt]
     0 & 0 & 0 & 0 & 0
    \end{bmatrix}\,.\quad
\]
Here, the matrix
$S=\left[\begin{smallmatrix} -\Bc & \\ \phantom{-}\Ac & -\Bc \\ & \phantom{-}\Ac \end{smallmatrix}\right]$
of size $12\times 10$ has rank $10$. Choosing the columns of $\Lc(\lambda)$ as $\Lc_i(\lambda)=\lb^0_i+\lb^1_i\lambda \in \CC^{5\times 1}$,  the linear systems
\begin{equation}
\label{SistemasEjemplo}
  \begin{bmatrix}
    -\Bc & \\ 
    \phantom{-}\Ac & -\Bc \\ 
        & \phantom{-}\Ac
  \end{bmatrix}
  \begin{bmatrix}
   \lb^1_i\\
   \lb^0_i
  \end{bmatrix}=
  \begin{bmatrix}
   \mathbf{0}\\
   \mathbf{0}\\
   I_4^i
  \end{bmatrix},\quad i=1,2,3,4
\end{equation}
have a unique solution. Observe that deleting the trivial equations 3 and 4, we have  consistent square systems.
By using  {\sf Matlab\texttrademark} we obtain the left inverse
\begin{multline*}
\Lc(\lambda)=10^3 \begin{bmatrix}
       \phantom{-}4.4812 &   -0.1438 &   \phantom{-}0.0166&   -0.0043\\
      -3.4840 &   \phantom{-}0.1118  & -0.0128 &   \phantom{-}0.0031\\
       \phantom{-}1.6069 &  -0.0514  &  \phantom{-}0.0056 &   \phantom{-}0.0000\\
      -0.4125  &  \phantom{-}0.0125  &  \phantom{-}0.0000 &  -0.0000\\
       \phantom{-}0.0500  &  \phantom{-}0.0000  & -0.0000 &   \phantom{-}0.0000
       \end{bmatrix}+ \\
       +10^3\begin{bmatrix}
        -0.0021 &   \phantom{-}0.0001 &  -0.0000 &   \phantom{-}0.0000\\
         \phantom{-}0.0517 &  -0.0017 &   \phantom{-}0.0002 &  -0.0000\\
         -0.4133 &   \phantom{-}0.0133 &  -0.0015 &   \phantom{-} 0.0004\\
         \phantom{-}1.6071  & -0.0516  &  \phantom{-}0.0059  & -0.0015\\
         -3.4841  &  \phantom{-}0.1118 &  -0.0129 &   \phantom{-}0.0033
       \end{bmatrix}\lambda
\end{multline*}

\medskip

At this point, the challenge problem is to give conditions on the matrix pencil $\Ac^\top-\lambda\Bc^\top$ in order to obtain a left inverse with polynomial entries (having nonegative powers) by solving the corresponding linear systems \eqref{SistemasEjemplo}. The answer to this question is based on the KCF of the matrix pencil $\Ac^\top-\lambda\Bc^\top$.
In our example the corresponding KCF is 
$N_1(\lambda)\oplus N_1(\lambda)\oplus L^\top_2(\lambda)$, i.e., the pencil has not finite eigenvalues, all the blocks associated with the infinite eigenvalue have order 1, and the left singular part has a unique block. In what follows, we prove that these conditions for the KCF of the matrix pencil $\widetilde{\GG}(\lambda)$ are sufficient to give a positive answer to the raised problem in a very important particular case:

\subsection{The case where the oversampling rate is minimum for a fixed $r\geq N$}

It corresponds to the case  where $N\leq r$ and $s=r+1$, i.e., the sampling period is 
$T=r/(r+1)$. Here, the matrix pencil $\widetilde{\GG}(\lambda)=\Ac^\top-\lambda\Bc^\top$ has the form
\begin{equation}
{\small
\label{formaHAZ}
\begin{bmatrix}
 0 &\cdots & 0 & 0 &\cdots  &  0 &  0  & 0\\
 0 &\cdots &0 & 0 &\cdots  & 0  & 0  & *\\
 0 &\cdots & 0 & 0 &\cdots  & 0 &*&*\\
 \vdots& & \vdots& \vdots&\iddots &  \iddots  & \vdots&\vdots\\
 0&\cdots &0 & 0&* &\cdots &* &*\\
 0& \cdots& 0&* &* &\cdots &* &*\\
 0& \cdots& *&* &* &\cdots &* &0\\
 \vdots& \iddots&  \vdots & \vdots & \vdots &\iddots & &\vdots\\
* & \cdots&* &* &*&\cdots&0& 0 
\end{bmatrix}-\lambda
\begin{bmatrix}
* &\cdots  & * &*&0   &  \cdots  & 0\\
 * &\cdots  &*  &*  & 0  & \cdots  & 0\\
 * &\dots  &*  & 0  & 0 & \cdots & 0\\
 \vdots&\iddots &\iddots &\vdots & \vdots   & &\vdots\\
  *&\iddots & 0 & 0&0 &\cdots &0\\
  0&\cdots &0 & 0&0 &\cdots & 0\\
0&\cdots &0& 0&0 &\cdots & 0\\
 \vdots& &\vdots &\vdots &\vdots & &\vdots\\
   0&\cdots  &0 &0 &0&\cdots& 0 
\end{bmatrix}\,,
}
\end{equation}
i.e., denoting the entries of $\Ac^\top$ and $\Bc^\top$ by $\Ac^\top_{ij}$ and $\Bc^\top_{ij}$ respectively,
we have $\Ac^\top_{ij}=0$ if $2+r<i+j<r+N+1$, $\Bc^\top_{1N}=0$ and $\Bc^\top_{ij}=0$ if $i+j>N+1$.
Having in mind the structure of the matrices $\Ac^\top$ and $\Bc^\top$ we have $\rango(\Ac^\top)\leq r$, 
$\rango(\Bc^\top)\leq N-1$ and 
$\rango(\left[\begin{smallmatrix} -\Bc & \\ \phantom{-}\Ac &\phantom{-} -\Bc \end{smallmatrix}\right])\leq r+N-1$. Whenever these matrices have maximum rank, the following result holds:

\begin{theorem}
\label{rangoG_s}
Assume that the singular matrix pencil $\Ac^\top-\lambda \Bc^\top$ of size $(r+1)\times r$ satisfies the following conditions:
\begin{enumerate}
 \item \label{condicion1} The pencil has not finite eigenvalues,
 \item $\rango(\Ac^\top)=r$,
 \item $\rango(\Bc^\top)=N-1$, with $N\leq r$, and
 \item $\rango(\left[\begin{smallmatrix}
                 -\Bc & \\
		\phantom{-}\Ac & \phantom{-}-\Bc 
               \end{smallmatrix}\right])=r+N-1$\,.
\end{enumerate}
Then, the $Nr\times(N-1)(r+1)$ matrix
\[
\Gc_r:=
\begin{bmatrix}
 -\Bc &   & & & &  \\
 \Ac &-\Bc & & & &\\
     &\Ac & -\Bc & & &\\
     &    &     &\ddots & &\\
     &    &     &       &\Ac & -\Bc\\
     &    &     &       &    &\Ac
\end{bmatrix}
\]
has rank $(N-1)(r+1)$.
\end{theorem}

First note that $\rango(\Ac^\top)=r$ implies that  the KCF of the matrix pencil $\Ac^\top-\lambda\Bc^\top$ has not right singular part (and also that 0 is not an eigenvalue). Thus, by using Theorem \ref{principal}, the pencil $\Ac^\top-\lambda\Bc^\top$ has a polynomial left inverse. Before to prove Theorem \ref{rangoG_s}, and in order to ease its proof, we first obtain, under the theorem hypotheses,  the KCF of the matrix pencil 
$\Ac^\top-\lambda\Bc^\top$: 
\begin{lemma}
\label{lema1}
The KCF of the matrix pencil $\Ac^\top-\lambda \Bc^\top$ is 
$\displaystyle{\left(\bigoplus_{i=1}^{r-N+1} N_1(\lambda)\right)\oplus L^\top_{N-1}(\lambda)}$.
\end{lemma}
{\bf Proof of Lemma \ref{lema1}: } Since the matrix pencil has neither finite eigenvalues nor right singular part, we conclude that its KCF has the form $N(\lambda)\oplus L^{left}(\lambda)$, where $N(\lambda)$ denotes the blocks associated with the infinite eigenvalue and $L^{left}(\lambda)$ denotes the left singular part. Since $r+1$ is the number of rows of the matrix pencil, $r$ the number of columns,  and the rank of $\Bc$ is $N-1$ it cannot appear blocks of the form  $L_i^\top(\lambda)$ for $i\geq N$. Each left singular block increases in one the number of rows with respect to the number of columns; hence, as the size of $\Ac^\top-\lambda \Bc^\top$ is  $(r+1)\times r$,  it can appear only one left singular block in its KCF. Furthermore, we prove that this only left singular block corresponds to $L_{N-1}^\top(\lambda)$. Indeed, let 
$\Kc_{\Ac}^\top-\lambda\Kc_\Bc^\top$ be the KCF of the matrix pencil $\Ac^\top-\lambda\Bc^\top$. Obviously, we have that 
$\rango(\Ac^\top)=\rango(\Kc_\Ac^\top)=r$,  $\rango(\Bc^\top)=\rango(\Kc_\Bc^\top)=N-1$ and
\[
\rango \begin{bmatrix}-\Bc & \\ \phantom{-}\Ac & -\Bc \end{bmatrix}=
\rango \begin{bmatrix}-\Kc_\Bc & \\ \phantom{-}\Kc_\Ac & -\Kc_\Bc \end{bmatrix}=r+N-1\,.
\]
The rank of the matrix $\left[ \begin{smallmatrix}-\Kc_\Bc & \\ \phantom{-}\Kc_\Ac & -\Kc_\Bc \end{smallmatrix}\right]$ coincides  with its number of nonzero rows because the number of null rows of 
$\Kc_\Bc$ is $r-N+1$, i.e., the number of blocks in $N(\lambda)$; the matrix $\Kc_\Ac$ has not null rows so that, the number of nonzero rows of 
$\left[ \begin{smallmatrix}-\Kc_\Bc & \\ \phantom{-}\Kc_\Ac & -\Kc_\Bc \end{smallmatrix}\right]$ is 
$2r-(r-N+1)=r+N-1$.

\medskip

Assume that in the KCF of the matrix  pencil $\Ac^\top-\lambda\Bc^\top$ appears a singular block 
$L_i^\top(\lambda)$ with $i< N-1$. Since the rank of $\Bc^\top$ is $N-1$, the regular part in the KCF has a block of the form $N_l(\lambda)$ with $l\geq 2$. By rearranging the blocks, we obtain that the KCF of
$\Ac^\top-\lambda\Bc^\top$ is $N_l(\lambda)\oplus\cdots\oplus L_i^\top(\lambda)$; therefore
\[
\begin{bmatrix}-\Kc_\Bc & \\ \phantom{-}\Kc_\Ac & -\Kc_\Bc \end{bmatrix}=
\begin{bmatrix}
\phantom{-}0 	& 0 & \cdots & 0 & 0 &0 & 0 &\cdots &0& 0\\
-1			& 0 & \cdots & 0 & 0 &0 & 0 &\cdots & 0&0\\
*	&*	&\cdots	&*	& *&0	&	0	& \cdots & 0& 0\\
\vdots & \vdots & & \vdots & \vdots & \vdots & \vdots & & \vdots & \vdots \\
*	&*	&\cdots	&*	& *&0	&	0	& \cdots & 0& 0\\
\phantom{-}1 	& 0 & \cdots & 0 & 0 &0 & 0 &\cdots &0& 0\\
 \phantom{-}0 	& 1 & \cdots & 0 & 0 &-1 & 0 &\cdots &0& 0\\
*	&*	&\cdots	&*	& *&*	&	*	& \cdots & *& *\\
\vdots & \vdots & & \vdots & \vdots & \vdots & \vdots & & \vdots & \vdots \\
*	&*	&\cdots	&*	& *&*	&	*	& \cdots & *& *\\
\end{bmatrix}
\]
In this case, the rank of 
$\left[ \begin{smallmatrix}-\Kc_\Bc & \\ \phantom{-}\Kc_\Ac & -\Kc_\Bc \end{smallmatrix}\right]$ is strictly smaller than $r+N-1$ because the second row and the $(r+1)$th row are linearly dependent. 
This contradicts the hypotheses and, hence, the only left singular block  is $L_{N-1}^\top(\lambda)$. Having in mind that $\rango(\Bc^\top)=N-1$, we conclude that the KCF of the matrix pencil $\Ac^\top-\lambda\Bc^\top$ is $\displaystyle{\left(\bigoplus_{i=1}^{r-N+1} N_1(\lambda)\right)\oplus L^\top_{N-1}(\lambda)}$.
{\hfill{$\square$}}

\medskip

{\bf Proof of Theorem \ref{rangoG_s}: } Once we have determined the KCF of the matrix pencil 
$\Ac^\top-\lambda\Bc^\top$ we compute the rank of the matrix $\Gc_r$. If $\Kc_{\Ac}-\lambda\Kc_\Bc$ is the KCF of the matrix pencil $\Ac-\lambda\Bc$, it is obvious that
\[
\rango(\Gc_r)=\rango \begin{bmatrix}
 -\Kc_\Bc &   & & & &  \\
 \phantom{-}\Kc_\Ac &-\Kc_\Bc & & & &\\
     &\phantom{-}\Kc_\Ac & -\Kc_\Bc & & &\\
     &    &     &\ddots & &\\
     &    &     &       &\phantom{-}\Kc_\Ac & -\Kc_\Bc\\
     &    &     &       &    &\phantom{-}\Kc_\Ac
\end{bmatrix}
\]
As $\Kc_\Ac^\top-\lambda \Kc_\Bc^\top$ is the KCF of the matrix pencil $\Ac^\top-\lambda \Bc^\top$, Lemma  \ref{lema1} gives
\[
  \Kc_\Ac^\top=\begin{bmatrix}
                 I 	& 0\\
		 0		& L_\Ac^\top
               \end{bmatrix},\quad
  \Kc_\Bc^\top=\begin{bmatrix}
                 0 	& 0\\
		 0		& L_\Bc^\top
               \end{bmatrix}
\] 
where $I=\mathbf{I}_{(r-N+1)}$ denotes the identity matrix of order $r-N+1$, and
\[
L_\Ac^\top=\begin{bmatrix}
            	0	& 0 		& \cdots	& 0\\
		1	& 0 		& \cdots	& 0\\
		\vdots	& \ddots	& \ddots	& \vdots\\		
		0	& 0 		& \cdots	&0\\
		0	& 0		&\cdots		&1
           \end{bmatrix}\in\CC^{N\times (N-1)}\,; \qquad
L_\Bc^\top=\begin{bmatrix}
            	1	& 0 		& \cdots	& 0\\
		0	& 1 		& \cdots	& 0\\
		\vdots	& \ddots	& \ddots	& \vdots\\		
		0	& 0 		& \cdots	&1\\
		0	& 0		&\cdots		&0
           \end{bmatrix}\in\CC^{N\times (N-1)}
\]
As a consequence,
\newcommand{\BigFig}[1]{\parbox{12pt}{\large #1}}
\newcommand{\BigZero}{\BigFig{0}}
\[
\rango(\Gc_r)=\rango
\begin{bmatrix}
 \BigZero & \phantom{-}\BigZero &  &  &  &  &  &  &  &  \\
 \BigZero & -L_\Bc &  &  &  &  &  &  &  &  \\
 I & \phantom{-}\BigZero & \phantom{-}\BigZero & \BigZero &  &  &  &  &  &  \\
 \BigZero & \phantom{-}L_\Ac & \phantom{-}\BigZero & -L_\Bc &  &  &  &  &  &  \\
  &  & I & \phantom{-}\BigZero & \BigZero & \BigZero &  &  &  &  \\
  &  & \phantom{-}\BigZero & \phantom{-}L_\Ac & \BigZero & -L_\Bc &  & &  &  \\
  &  &  &  & \ddots &  & \ddots &  &  &  \\
  &  &  &  &  &  & I & \phantom{-}\BigZero & \BigZero & \BigZero \\
  &  &  &  &  &  & \BigZero & \phantom{-}L_\Ac & \BigZero & -L_\Bc \\
  &  &  &  &  &  &  & & I & \phantom{-}\BigZero \\
  &  &  &  &  &  &  & & \phantom{-}\BigZero & \phantom{-}L_\Ac 
\end{bmatrix}
\]
A suitable interchange of rows and columns gives
\[
\rango(\Gc_r)=\rango
\begin{bmatrix}
 \BigZero & \cdots & \BigZero &  &  &\phantom{-}\BigZero  &\phantom{-}\BigZero  \\
 I &  &  &  &  &  &  \\
  & \ddots &  &  &  &  &  \\
  &  & I &  &  &  &  \\
  &  &  & -L_\Bc & 0 &  &  \\
  &  &  & \phantom{-}L_\Ac &-L_\Bc  &  &  \\
  &  &  &  &  & \ddots &  \\
  &  &  &  &  & \phantom{-}L_\Ac  &-L_\Bc \\
  &  &  &  &  & 0 &\phantom{-}L_\Ac 
\end{bmatrix}
\] 
where the first $r-N+1=r-\rango(\Kc_\Bc)$ are null rows; hence, the rank of $\Gc_r$ equals $(N-1)(r+1)$ if and only if the remaining $(N-1)(r+1)$ rows are linearly independent. This is equivalent to the matrix
\[
\Ll_{\Ac,\Bc}=\begin{bmatrix}
 -L_\Bc & \phantom{-}\BigZero &  &  \\
  \phantom{-}L_\Ac &-L_\Bc  &  &  \\
  &  &   \ddots &  &\\
  &  &  \phantom{-}L_\Ac&  -L_\Bc  \\
  &  & \phantom{-}\BigZero &  \phantom{-}L_\Ac 
\end{bmatrix}\in\CC^{N(N-1)\times N(N-1)}
\]
has full rank. To prove it, we use the following  result in \cite[p. 32]{gantmacher:00}: Let $\xb(\lambda)$ be a nonzero vector having the form 
\[
 \xb(\lambda)=\xb_0+\lambda\xb_1+\lambda^2\xb_2+\cdots+\lambda^\varepsilon \xb_\varepsilon, 
\,\, \xb_i\in \CC^{N\times 1}
\]
such that $(L_\Ac-\lambda L_\Bc)\xb(\lambda)=\mathbf{0}$.  Then, necessarily, $\varepsilon\geq N-1$.
Now, let us continue by contradiction, and assume  that  the matrix $\Ll_{\Ac,\Bc}$ has not full rank. Then, there exists a nonzero vector $\mathbf{z}\in \CC^{N(N-1)\times 1}$
such that $\Ll_{\Ac,\Bc}\,\mb{z}=\mb{0}$.  Denoting 
$\mb{z}^\top=[\mb{z}_{N-2}^\top\, \dots\,\mb{z}_1^\top\, \mb{z}_0^\top]$ where 
$\mb{z}_i\in \CC^{N\times 1}$,
we obtain that
\[
(L_\Ac-\lambda L_\Bc)(\mb{z}_0+\lambda\mb{z}_1+\lambda\mb{z}_2+\cdots +\lambda^{N-2}\mb{z}_{N-2})=\mb{0},
\]
which contradicts the minimal property for $N-1$. Therefore, the matrix $\Ll_{\Ac,\Bc}$ has full rank
and, finally,  $\rango\Gc_r=(N-1)(r+1)$.
{\hfill{$\square$}}

\medskip

\noindent {\bf Remark:} Note that Theorem \ref{rangoG_s} remains valid for any singular matrix pencil 
$\Ac^\top-\lambda\Bc^\top$ of size $(r+1)\times r$ substituting $N-1$ by $p\in \NN$ which satisfies $0<p<r$.

\medskip

Consider the matrix pencil  $\widetilde{\GG}(\lambda)=\Ac^\top-\lambda\Bc^\top$ of size $(r+1)\times r$ with $N\leq r$. Assuming that the $\widetilde{\GG}(\lambda)$ has polynomial left inverses,  the following result gives sufficient conditions for computing one of such polynomial left inverses. Once we have got one solution, it is straightforward to derive the remaining solutions.

\begin{corollary}[Computing a polynomial left inverse of $\widetilde{\GG}(\lambda)$]
\label{calculoInvIzq}
Let $\widetilde{\GG}(\lambda)=\Ac^\top-\lambda \Bc^\top$ be a singular matrix pencil of size  
$(r+1)\times r$ with $N\leq r$.  Assume that $\widetilde{\GG}(\lambda)$ admits polynomial left inverses, and that the following conditions hold:
\begin{enumerate}
\item\label{1} $\rango(\Ac^\top)=r$,
\item\label{2} $\rango(\Bc^\top)=N-1$,  and
\item\label{3} $\rango(\left[\begin{smallmatrix}
                 -\Bc & \\
		\phantom{-}\Ac & \phantom{-}-\Bc 
               \end{smallmatrix}\right])=r+N-1$
\end{enumerate}
Then, the linear systems
\begin{equation}
\label{sistemaG_s}
\begin{bmatrix}
 -\Bc &   & & & &  \\
 \Ac &-\Bc & & & &\\
     &\Ac & -\Bc & & &\\
     &    &     &\ddots & &\\
     &    &     &       &\Ac & -\Bc\\
     &    &     &       &    &\Ac
\end{bmatrix}
\begin{bmatrix}
 \lb_i^{N-2}\\
  \lb_i^{N-3}\\
    \vdots\\
 \lb_i^{0\phantom{-1}}\\
\end{bmatrix}=
\begin{bmatrix}
 \mathbf{0}\\
  \vdots\\
 \mathbf{0}\\
    \vdots\\
    \mathbf{0}\\
      I_r^i\\
\end{bmatrix},\quad i=1,2,\dots,r\,,
\end{equation}
where $I^i_r$ denotes the $i$th column of the identity matrix $\mathbf{I}_r$, admit a unique solution. Moreover,  let
$[\lb_i^{N-2}\,\lb_i^{N-3}\,\dots\,\lb_i^{0}]^\top \in \CC^{(N-1)(r+1)}$ be this solution for $i=1,2,\ldots,r$, and consider 
the polynomial vector $\Lc_i(\lambda)=\lb^0_i+\lb_i^1 \lambda+\cdots+\lb_i^{N-2}\lambda^{N-2}$, $i=1,2,\dots,r$. Then, the $(r+1)\times r$ polynomial matrix
\[
 \Lc(\lambda):=[\Lc_1(\lambda)\, \Lc_2(\lambda)\, \dots \, \Lc_r(\lambda)]
\]
satisfies
\[
\Lc^\top(\lambda)\widetilde{\GG}(\lambda)=\mathbf{I}_r
\] 
\end{corollary}
{\bf Proof: } Theorem \ref{rangoG_s} implies that the rank of the coefficient matrix  
$\Gc_r \in\CC^{Nr\times(N-1)(r+1)}$ is $(N-1)(r+1)$ in  \eqref{sistemaG_s}.
Having in mind \eqref{formaHAZ}, the last $r-N+1$ rows of $\Bc$ are null. Deleting these rows in the first row block (which become trivial equations in \eqref{sistemaG_s}), we obtain an square invertible matrix, and consequently \eqref{sistemaG_s} has a unique solution for each $i=1,2,\ldots,r$. Recalling \eqref{matAB2}, we finally obtain that $\Lc^\top(\lambda)$ is a polynomial left inverse of  $\widetilde{\GG}(\lambda)$.
{\hfill{$\square$}}

\medskip

Observe that any other polynomial left inverse $A(\lambda)$ of the matrix $\widetilde{\GG}(\lambda)$ is given by
\[
A(\lambda)=\Lc^\top(\lambda)+B(\lambda)\big[\mathbf{I}_{r+1}-\widetilde{\GG}(\lambda)\Lc^\top(\lambda)\big]\,,
\]
where $B(\lambda)$ is an arbitrary $r\times (r+1)$ polynomial matrix.

\medskip

For the matrix pencil $\widetilde{\GG}(\lambda)=\Ac^\top-\lambda \Bc^\top$ of size $(r+1)\times r$ with 
$N\leq r$, it is easy to give sufficient conditions in order to satisfy the conditions \ref{1}-\ref{3} in 
Corollary \ref{calculoInvIzq}. Namely:
\begin{corollary}
Consider the singular matrix pencil $\widetilde{\GG}(\lambda)=\Ac^\top-\lambda \Bc^\top$ of size 
$(r+1)\times r$ with $N\leq r$. Denoting $\Ac^\top=[\Ac^\top_{ij}]$ and 
$\Bc^\top=[\Bc^\top_{ij}]$, assume that the following conditions hold:
\begin{align}
\label{condSuf1}
 \Ac^\top_{ij}&\neq 0\,\text{ if $i+j=r+2$ or  $i+j=r+N+1$}\\
\label{condSuf2}
 \Bc^\top_{ij}&\neq 0\,\text{ if $i+j=N+1$ and $i\geq 2$}\,.
\end{align}
Then the conditions \ref{1}-\ref{3} in Corollary \ref{calculoInvIzq} are satisfied.
\end{corollary}
{\bf Proof: } Conditions \eqref{condSuf1} and \eqref{condSuf2} say that the entries marked as $\bullet$ in the matrices below are nonzero
\begin{align*}
\setcounter{MaxMatrixCols}{13}
\Ac&={\small\left[\begin{array}{cccccccc|cccc}
 0 & \hdotsfor{3}&0 & 0 &\cdots & \cdots& 0 & \cdots & 0 & \bullet \\
 0 & \hdotsfor{3}&0 & 0&\cdots & \cdots & 0 & \cdots & \bullet & * \\
 \vdots & & &  &\vdots  & \vdots &  & \vdots &\vdots  & \iddots & \vdots & \vdots \\
 \vdots & & &  &\vdots  & \vdots &  & \vdots  &\iddots  &  & \vdots & \vdots \\
 \vdots & && &\vdots &\vdots & & \iddots&\vdots & &\vdots & \vdots\\
 0 &\hdotsfor{3} & 0 & 0 & \bullet & \cdots & * & \cdots & * &*  \\
 0 &\hdotsfor{3} & 0 & \bullet & * & \cdots&  *& \cdots & * &*  \\
\hline
 0 & \hdotsfor{2} & 0 & \bullet & * & * & \cdots & *& \cdots & *&\bullet  \\
 0 & \hdotsfor{2}&  \bullet&  *& * & * & \cdots & * & \cdots & \bullet & 0 \\
 \vdots & &\iddots &  &  &  &  &  \vdots& \vdots &\iddots  &  &\vdots  \\
 0 & \bullet &\cdots &*  & * & * &\cdots  &  \cdots& \bullet & \cdots & 0 &0 
    \end{array}\right]}=\begin{bmatrix}
    A_{11} & A_{12}\\
   A_{21}& A_{22}
   \end{bmatrix}\in\CC^{r\times(r+1)}\\
 & \\
\Bc&={\small\left[\begin{array}{ccccc|ccc}
  *&*  & \cdots & * & \bullet &0  & \cdots  &0  \\
  *&*  &\cdots  & \bullet & 0 &0  &\cdots  &0  \\
  \vdots&\vdots  &\iddots  &\vdots  &\vdots  &\vdots  &  & \vdots \\
  *& \bullet &\cdots  & 0 &0  &0  & \cdots & 0 \\
  \hline
  0& 0 &\cdots  & 0 &0  &0  & \cdots & 0\\
  \vdots& \vdots &  &\vdots  &\vdots  &\vdots  &  &\vdots  \\
  0& 0 &\cdots  & 0 &0  &0  & \cdots & 0
     \end{array}\right]}=\begin{bmatrix}
    				B_{11} & 0\\
				0	       & 0
   			 \end{bmatrix}\in\CC^{r\times(r+1)}
\end{align*}
where $A_{22}\in\CC^{(r-N+1)\times(r-N+1)}$ and $B_{11}\in\CC^{(N-1)\times N}$. Trivially, 
$\rango(\Ac^\top)=r$ and $\rango(\Bc^\top)=N-1$.
Condition \ref{3} comes by observing the form of the matrix $\left[\begin{smallmatrix} -\Bc & \\ \phantom{-}\Ac & \phantom{-}-\Bc \end{smallmatrix}\right]$.
Interchanging rows and columns we obtain that the matrix 
$\left[\begin{smallmatrix} -\Bc & \\ \phantom{-}\Ac & \phantom{-}-\Bc \end{smallmatrix}\right]$ has the same rank than the matrix
\[
\begin{bmatrix}
 0	& 0 & 0 &0\\
 B_{11} & 0 & 0 & 0\\
 A_{11}      &B_{11} & A_{12} &0\\
 A_{21} & 0     & A_{22} & 0
\end{bmatrix}
\]
Since the matrix $A_{22}\in\CC^{(r-N+1)\times(r-N+1)}$  is invertible, elementary row operations give the new matrix
\[
\begin{bmatrix}
 0	& 0 & 0 &0\\
 B_{11} & 0 & 0 & 0\\
 \widetilde{A}_{11}      &B_{11} & 0 &0\\
 A_{21} & 0     & A_{22} & 0
\end{bmatrix}\,.
\]
Finally, the above matrix has rank $2(N-1)+r-N+1=r+N-1$.

{\hfill{$\square$}}

\medskip

Remark that condition \ref{condicion1} in Theorem \ref{rangoG_s} can be checked by using the algorithm
{\sf guptri}. In case that conditions \ref{1}-\ref{3} in Corollary \ref{calculoInvIzq} are satisfied, we could check directly the consistency of the linear systems \eqref{sistemaG_s};  if they are not consistent, we derive that the pencil $\widetilde{\GG}(\lambda)$ has not polynomial left inverses.

\section{Conclusion}

Consider the problem of the recovery of any function $f$ in a shift-invariant space $V_\varphi$ from the sequence of samples $\{\big(\mathcal{L}f\big)(rn/s)\}_{n\in \ZZ}$ of its filtered version 
$\mathcal{L}f$, where the positive integers $r$ and $s$ satisfy $s>r$, i.e., the oversampling setting.
The existence of compactly supported reconstruction functions for this sampling problem is intimately related to the existence of a polynomial left inverse for a polynomial matrix $\mathsf{G}(z)$ associated with the sampling problem. This is equivalent to that the matrix 
$\mathsf{G}(z)$ has full rank for any $z \in \CC\setminus\{0\}$.  Other characterizations can be found in the mathematical literature involving the Smith canonical form of the polynomial matrix $\mathsf{G}(z)$
or the Euclides algorithm for the minors of order  $r$ in $\mathsf{G}(z)$. Unfortunately, whenever the parameter $r$ is large,  the aforesaid methods are useless from a practical point of view. In this work, by assuming that  $N\leq r < s$, where $\mathcal{L}\varphi \subseteq [0,N]$, we derive a new characterization for the existence of polynomial left inverses which involves the Kronecker canonical form of a singular matrix pencil. The advantage, from a practical point of view, of this new method is that
we can retrieve the needed information from the KCF by using the so-called {\sf guptri} algorithm. 
Furthermore, in the  important case where $s=r+1$, i.e.,  the oversampling rate is minimum (for a fixed 
$r$), we propose a method for the computation of a polynomial left inverse (and hence, the whole set of polynomial left inverses) of $\mathsf{G}(z)$.

\section{Appendix}

For the sake of completeness we include here a brief reminder on the canonical forms alluded throughout
the paper.

\subsection{Smith canonical form of a polynomial matrix}
\label{smith}

Recall that any $m\times n$ ($m\geq n$) polynomial matrix $\mathsf{H}(z)$ with 
$\rango \mathsf{H}(z)=r$ (recall that the rank of a polynomial matrix is the order of its largest minor that is not equal to the zero polynomial)  can be written as the product  
$\mathsf{H}(z)=\mathbf{V}(z)\mathsf{S}(z)\mathbf{W}(z)$
where $\mathbf{V}(z)$ an $\mathbf{W}(z)$ are unimodular matrices 
(i.e., the determinants of $\mathbf{V}(z)$ and $\mathbf{W}(z)$ are nonzero constants)
of size $m\times m$ and $n\times n$ respectively, and $\mathsf{S}(z)$ is a diagonal $m\times n$ polynomial matrix $\mathsf{S}(z):=\mbox{diag}[i_1(z),\ldots,i_r(z),0,\ldots,0]$. Moreover, the diagonal entries (the so-called invariant polynomials of $\mathsf{H}(z)$) are given by
$i_j(z)=d_j(z)/d_{j-1}(z)$, $j=1,2,\ldots,r$, where $d_j(z)$ is the greatest common divisor of all minors of 
$\mathsf{H}(z)$ of order $j$, $j=1,2,\ldots,r$, and $d_0(z)\equiv 1$. The matrix $\mathsf{S}(z)$ is called the Smith canonical form of the matrix $\mathsf{H}(z)$. See \cite{lancaster:85} for the details.

\subsection{Kronecker canonical form of a matrix pencil}
\label{FCK}

The Kronecker canonical form (KCF) for matrix pencils $\HH(\lambda)=A-\lambda B$, 
$A,B\in \CC^{m\times n}$, is a generalization of the Jordan canonical form to matrix pencils (see 
\cite{gantmacher:00}): There exist two nonsingular matrices $U\in\CC^{m\times m}$ and $V\in\CC^{n\times n}$ such that (in block structure notation):
\[
U(A - \lambda B)V^{-1}= S^{right}_\HH(\lambda) \oplus J_\HH(\lambda) \oplus N_\HH(\lambda)\oplus 
S^{left}_\HH(\lambda)\,,
\]
where $J_\HH(\lambda) \oplus N_\HH(\lambda)$ is the regular part of the matrix pencil, 
$S^{right}_\HH(\lambda)$ is its right singular part, and $S^{left}_\HH(\lambda)$ its left singular part.
The block $J_\HH(\lambda)$ is associated with the finite eigenvalues of the matrix pencil, and it reads:
\[
   J_\HH(\lambda)=J_{l_1}(\mu_1)\oplus\cdots \oplus J_{l_{g_q}}(\mu_q)
\]
where $J_{lj}(\mu_i )$ is a  $l_j\times l_j$ Jordan block associate with the finite eigenvalue $\mu_i$, i.e.,
\[
J_{l_i}(\mu_i)=\begin{bmatrix}
               \mu_i & 1 & & \\
                     & \ddots & \ddots & \\
                     &        & \ddots  & 1\\
                     &        &        &\mu_i
              \end{bmatrix}-\lambda \begin{bmatrix}
               				 1 & 0 & & \\
                     			      & \ddots & \ddots & \\
                     			      &        & \ddots  & 0\\
                     			      &        &        & 1
              \end{bmatrix}
\]
Recall that $\mu$ is a finite eigenvalue of the matrix pencil $\HH(\lambda)$ if 
$\rango \HH(\mu) < \rango \HH(\lambda)$, being $\rango \HH(\lambda)$ the order of the largest minor that is not equal to the zero polynomial.

The block $N_\HH(\lambda)$ is associated with the infinite eigenvalue (if does exist) and it reads:
 \[
   N_\HH(\lambda)=N_{p_1}(\lambda)\oplus\cdots \oplus N_{p_{g_\infty}}(\lambda)
\]
where $N_{p_i}\in \CC^{p_i\times p_i}$ is the matrix
\[ 
N_{p_i}(\lambda)=\begin{bmatrix}
               1 & 0 & & \\
                     & \ddots & \ddots & \\
                     &        & \ddots  & 0\\
                     &        &        &1
              \end{bmatrix}-\lambda \begin{bmatrix}
               				 0 & 1 & & \\
                     			      & \ddots & \ddots & \\
                     			      &        & \ddots  & 1\\
                     			      &        &        & 0
              \end{bmatrix}
\]
and $g_\infty$ denotes the geometric multiplicity of the infinite eigenvalue which corresponds  to the number of Jordan blocks for the infinite eigenvalue. Recall that the pencil $\HH(\lambda)$ has the infinite eigenvalue if its dual pencil $\HH^\sharp(\lambda):=\lambda \HH(1/\lambda)$ has the zero eigenvalue.

If $m\neq n$ or $\det(A-\lambda B)=0$ for all $\lambda \in \CC$, then the matrix pencil also includes a singular part,  $S^{right}_\HH(\lambda)$ and/or $S^{left}_\HH(\lambda)$, and we say that the matrix pencil is singular. For the right singular part, we have
\[
S^{right}_\HH(\lambda)=L_{\varepsilon_1}(\lambda)\oplus \cdots \oplus L_{\varepsilon_{r_0}}(\lambda)
\]
where $L_{\varepsilon_i}(\lambda)$ is a block of size $\varepsilon_i\times(\varepsilon_i + 1)$ defined by
\begin{equation}
\label{LS}
L_{\varepsilon_i}(\lambda)=\begin{bmatrix}
                 0 & 1 & & \\
                     & \ddots & \ddots & \\
                     &        &      0  &1
              \end{bmatrix}-\lambda \begin{bmatrix}
               				 1 & 0 & & \\
                     			      & \ddots & \ddots & \\
                     			      &        &    1    & 0
              \end{bmatrix}\,.
\end{equation}
$L_0$ is a block of size $0 \times 1$ which contributes to a column of zeros. Analogously, the left singular part has the form
\[
S^{left}_\HH(\lambda)=L^T_{\eta_1}(\lambda)\oplus \cdots \oplus L^T_{\eta_{l_0}}(\lambda)\,,
\]
where $L_{\eta_i}^\top(\lambda)$ is a block of size $(\eta_i + 1)\times \eta_i$, and $L^\top_0$ is a block of size $1 \times 0$ which contributes to a row of zeros.

\subsection{GUPTRI form}
\label{formaguptri}

The GUPTRI form (Generalized UPer TRIangular form) for singular matrix pencils was done by Van Dooren \cite{Dooren:97,Dooren:98} by using unitary equivalence transformations. It is a generalization of the Schur-staircase form for matrices.

Given  a singular matrix pencil $\HH(\lambda)=A-\lambda B$ with $A,B\in\CC^{m\times n}$, there exist two unitary matrices $U$ and $V$ of size  $m\times m$ and  $n\times n$ respectively such that
\[
  U(A-\lambda B)V^H=\begin{bmatrix}
              A_r-\lambda B_r 	& \star & \star\\
		0	& A_{reg}-\lambda B_{reg} & \star\\
		0	& 0		   & A_l-\lambda B_l
             \end{bmatrix}\,,
\]
where the rectangular block upper triangular $A_r-\lambda B_r$ and $A_l-\lambda B_l$ give the right and left singular structures of the matrix pencil, respectively. The remaining square upper triangular 
$A_{reg}-\lambda B_{reg}$ contains all the finite and infinite eigenvalues of $\HH(\lambda)$. Furthermore, the regular part  $A_{reg}-\lambda B_{reg}$ is in staircase form:
\[
A_{reg}=\begin{bmatrix}
         A_0 & \star & \star \\
          0  & A_f &\star\\
          0  & 0   & A_\infty
        \end{bmatrix},\quad
B_{reg}=\begin{bmatrix}
         B_0 & \star & \star \\
          0  & B_f &\star\\
          0  & 0   & B_\infty
        \end{bmatrix}
\]
where $A_0-\lambda B_0$ and $A_\infty-\lambda B_\infty$ reveal the Jordan structures of the zero and infinite eigenvalues, and $A_f-\lambda B_f$, in generalized Schur form, includes the finite but nonzero eigenvalues.

The eigenvalues $\mu_i$ are computed as pairs of values, denoted by $(\alpha_i,\beta_i)$, $\alpha_i$ in the diagonal of $A_{reg}$ and $\beta_i$ in the diagonal of $B_{reg}$ as follows: If $\alpha_i \neq 0$ and $\beta_i\neq  0$ then $\mu_i$ is the finite nonzero eigenvalue $\mu_i=\alpha_i/\beta_i$; if 
$\alpha_i=0$ and $\beta_i\neq  0$, $\mu_i$ is the zero eigenvalue and; if $\alpha_i \neq 0$ and 
$\beta_i=0$ then $\mu_i$ is the infinite eigenvalue.  The case $\alpha_i=0$ and  $\beta_i=0$ does not correspond to an eigenvalue, instead it belongs to the singular part of the matrix pencil.
In \cite{Demmel:93a, Demmel:93b} is described an efficient algorithm for computing the GUPTRI form
of a matrix pencil. The implementation of this algorithm can be found in  {\verb http://www.cs.umu.se/research/nla/singular_pairs/guptri }.

\bigskip

\noindent{\bf Acknowledgments:} 
The authors gratefully acknowledge F. M. Dopico (Universidad Carlos III de Madrid) for the fruitful discussions on the proof of Theorem \ref{rangoG_s}.
This work has been supported by the grant MTM2006--09737
from the D.G.I. of the Spanish {\em Ministerio de Ciencia y Tecnolog\'{\i}a}.

\end{document}